\documentclass{optica-article}
\usepackage[utf8]{inputenc}
\usepackage{mathtools}
\usepackage{amsmath}
\usepackage{physics}
\usepackage{graphicx}

\usepackage{color}
\usepackage{graphicx}
\usepackage{lineno} 
\usepackage{epstopdf}%

\newcommand {\be} {\begin{equation}} 
\newcommand {\ba} {\begin{eqnarray}} 
\newcommand {\ee} {\end{equation}} 
\newcommand{\ea} {\end{eqnarray}}

\renewcommand{\epsilon}{\varepsilon}

\journal{optica}

\usepackage{lineno}

\begin{document}

\title{Non-Diffracting Polarisation Features around Far-Field Zeros of Electromagnetic Radiation}
\author{Alex J. Vernon,\authormark{1} Andrew Kille,\authormark{2}, Francisco J. Rodr\'iguez-Fortu\~no\authormark{1, 3,*}, and Andrei Afanasev\authormark{2, 4,*}}

\address{\authormark{1}Department of Physics and London Centre for Nanotechnology, King's College London, Strand, London WC2R 2LS, UK\\
\authormark{2}Department of Physics, The George Washington University, Washington, DC 20052, USA}
\authormark{3}francisco.rodriguez\_fortuno@kcl.ac.uk\\
\authormark{4}afanas@gwu.edu\\
\authormark{*}Corresponding authors

\begin{abstract}
Light from any physical source diffracts over space, as spherical wavefronts grow and energy density is spread out.
Diffractive effects pose fundamental limits to light-based technologies, including communications, spectroscopy, and metrology.
Polarisation becomes paraxial in the far field limit and, by ignoring longitudinal field components, the rich physics of non-paraxial fields which exist in near-fields or a beam's tight focus are lost.
The longitudinal field cannot, however, be ignored when transverse field components vanish (in a transverse field zero) and carry a small non-paraxial region to infinity.
We show that a transverse field zero is always accompanied by non-diffracting polarisation structures, whose geometries are independent of the distance to the source, including an enclosing intensity ratio tube, and parallel, non-diverging polarisation singularities.
We illustrate these features in multipole radiation and in double slit interference, two examples which have time-fixed transverse field zeros.
Non-diffracting structures with changing position are coupled to time-varying zeros, which are present in all far field radiation.
\end{abstract}


\section{Introduction}	\label{sec:intro}

The energy of a propagating electromagnetic wave spreads out in space, inevitably, which is to say that light diffracts as it propagates from any localised source.
Any theoretical exception to this rule, such as a Bessel beam, cannot be generated by a physically realisable source.
Travelling a sufficiently far distance from a localised source of size $D$, that is $r\gg D$, light falls under the paraxial approximation as its polarisation, spin angular momentum and linear momentum become constrained and nearly homogeneous over wavelength scales.
Longitudinal (radial) electric and magnetic field components are overwhelmed by the transverse components ($\boldsymbol{\hat{\theta}}$ and $\boldsymbol{\hat{\phi}}$), polarising the electromagnetic field almost perpendicular to the wavevector.
The diminished radial field component can neither tilt the polarisation ellipse enough to develop a noticeable transverse spin component, nor impart a significant transverse component to the Poynting vector, which also points radially, parallel to the wavevector.

A much richer physics is found in non-paraxial electromagnetic fields, such as tightly focused beams and near fields.
Extraordinary properties of light and structures emerge here, including dominating transverse spin \cite{Bliokh2014,Banzer2013,Bekshaev2015,Eismann2021}, superoscillations \cite{Dennis2008superoscillations,Berry1994,Rogers2013}, skyrmions \cite{Tsesses2018,Sugic2021}, and topological momentum structures \cite{Kleckner2013,Berry2019,Leach2004, Vernon2023}, and counter-intuitive spin features in the near field of even the simplest of dipolar sources \cite{Neugebauer2019}.
Many of these features are naturally stirred into the light field by phase and polarisation singularities, such as C lines, threads of circular polarisation, and L lines, threads of linear polarisation \cite{nye1983lines,Nye1987,Berry2001,berry2004electric}.
These and other one-dimensional singularities may organise closed, sometimes knotted, loops \cite{Larocque2018}, or escape the non-paraxial field, propagating infinitely into the far field and never terminating.

Recently, a new polarisation structure was defined around the vortex centre of a doughnut beam \cite{afanasev2023nondiffractive}.
The centre of the beam is, in fact, not completely dark due to a persisting, small longitudinal field component.
We might imagine the vortex centre as a highly symmetric L line, longitudinally polarised, as both electric field components transverse to the beam direction are zero.
This type of zero creates an enclosing, uniform tube, where the intensity of the transverse components is either equal or proportional by a constant factor to the radial component intensity.
Surprisingly, this tube does not diffract, even at an infinite distance from the focus of the beam.
Though it becomes less obvious away from the focus, the central L line of the beam extends infinitely in the longitudinal direction and preserves the cross section of the tube.
A non-diffracting tube of circular polarisation was also recently predicted around the axis of a radiating dipole \cite{Mok2022}.

In this work, we reveal that these non-diffracting polarisation features are not limited to the singularity of a doughnut beam or a dipole's axis, but are rather a completely general phenomenon and a direct consequence of Maxwell's equations, existing around any far field transverse polarisation zero emanating from any arbitrary localised source.
We show that this type of zero carries polarisation and momentum structures, characteristic of non-paraxial light, well into the far field, sustained at arbitrarily large distances from the source.
These structures include non-diffracting intensity tubes and non-diverging, parallel C lines.

One may consider time-(in)dependent zeros using either the complex transverse field phasor or real instantaneous vector.
Time-varying transverse field zeros are stable polarisation singularities and at least one \textit{must always} exist in far field radiation, regardless of the source.
Non-diffracting structures with time-dependent position, therefore, are ubiquitous phenomena in physical propagating fields.
Non-diffracting designed beams whose intensity profile is invariant to propagation, such as Bessel beams, exist in theory, but in practice their non-diffractive behavior is limited to distances comparable to the source spatial extent.
The non-diffracting polarisation features we present in this work are very different in nature, as they exist in the polarisation structure of the fields, superimposed on top of a normally-diffracting intensity profile, and they exist for any source with a far field zero, regardless of size and distance.
Unlike those of an optical vortex \cite{afanasev2023nondiffractive}, the non-diffracting and near field-like objects present in general far field radiation could be realistically measured on any frequency range, for example the microwave band, using conventional radiating antennas.
If measured, a non-diffracting object could be an effective tool for beam alignment in metrology, or in RF, microwave, and terahertz communications.
In these centimetre and millimetre regimes, the non-diffracting object has a human-appreciable size. 
The use of zeros for alignment is known in radio frequency, where radiation diagrams having sharp zeros allows finding the direction of a signal's origin by aligning it to the antenna's null. Still, locating a zero with precision is challenging in the presence of noise - the existence of polarisation structures forming tubes around the zeros suggest robust methods of alignment, where if the small non-diffracting tube is detected, it is guaranteed to contain the zero in its small interior.

We begin in section \ref{sec:one} with a general formalism of the non-paraxial and non-diffracting features created by a transverse field zero in any kind of far field radiation.
In section \ref{sec:two}, we focus on specific examples of multipole radiation, which creates special symmetric non-diffracting objects and higher dimensional polarisation structures.
The widespread presence of non-diffracting structures is demonstrated in section \ref{sec:doubleslit}, appearing even in Young's centuries-old double slit experiment.
Finally, we conclude in section \ref{sec:conclusions}.


\section{General formalism}			\label{sec:one}

\begin{figure}[t]
    \centering
    \includegraphics[width=0.5\textwidth]{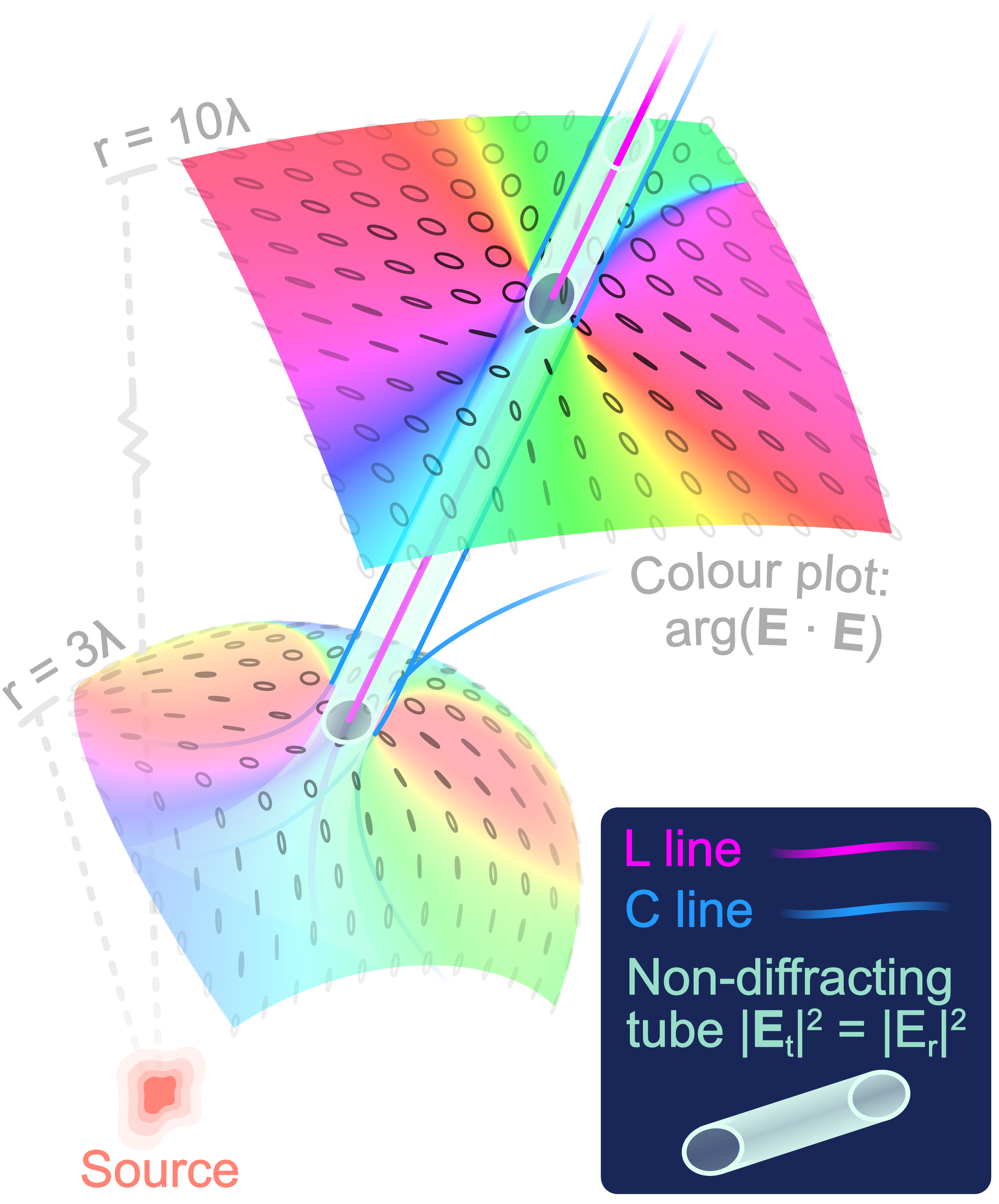}
    \caption{Non-diffracting objects in arbitrary radiation from a source, containing a zero in the transverse field components (indicated by the magenta line, a radially polarised L line).
    The surface connecting all real space points where the transverse field magnitude equals the radial component magnitude, $|E_\theta|^2+|E_\phi|^2=|E_r|^2$, is a non-diffracting tube with an elliptical cross section which is preserved with radial distance from the source.
    Two parallel C lines (blue lines) accompany the tube into the far field.
    On patches of two spheres, one of radius $r=3\lambda$ and the other $r=10\lambda$, the phase angle of the complex scalar field $\mathbf{E\cdot E}$ is plotted (C lines are singularities in this scalar field), as well as normalised 3D polarisation ellipses.
    Non-diffracting features exist in part because of the approximately linear behaviour of the transverse field with respect to $(\theta,\phi)$ in the neighbourhood of the zero.
    Near to the source (beneath the $r=3\lambda$ patch), the tube and polarisation singularities dissociate as higher-order behaviour of $\mathbf{E}$ becomes significant relative to the transverse extent of the tube (hence why the tube is not quite elliptical in cross section on the $r=3\lambda$ patch).
    A third C line briefly escorts the tube at $r=3\lambda$, then diverges for increasing $r$ as $\mathbf{E}$ becomes more linear in the region of the zero, making $\mathbf{E\cdot E}$ more symmetric and coaxing the number of accompanying C lines towards an even number.}
    \label{fig1}
\end{figure}
The far field radiation from any localised source can be described by a radiation diagram which contains the polarisation information of the transverse electric field components, $E_\theta$ and $E_\phi$.
The diagram neglects the small radial electric field component, insignificant unless, crucially, $E_\theta$ and $E_\phi$ are zero simultaneously.
This transverse field zero establishes a small non-paraxial region where the longitudinal component can no longer be ignored, dominating, and organising a radially polarised L line in the zero's position.
Light in this neighbourhood is still very dim, however, and as the transverse field grows to a competing magnitude with the radial component, typical non-paraxial features such as one-dimensional C lines and transverse spin appear.

Two non-diffracting objects are illustrated in Fig. \ref{fig1} around a transverse zero (the radially polarised L line), one: a non-diffracting tube of constant cross section, extending radially from a source, connecting all real-space points where the transverse field and radial component intensities are equal, and two: C lines, lines of pure circular polarisation, which remain parallel to each other, near to the tube, as they propagate in the radial direction.
We begin by defining the broad mathematical criteria that, on the surface of a sphere, draw a closed loop whose area is preserved even as the sphere increases in radius.
We will show that a comparison of the far field intensity of different components of light from any localised source can fulfil these criteria.

A line on the surface of a sphere of unspecified radius takes the form,
\begin{equation}\label{eq1}
    F(\theta,\phi)=\delta,
\end{equation}
\noindent where $(\theta,\phi)$ are the polar angles and $\delta$ is a constant.
This single condition generically defines a one-dimensional line because the parameter space $(\theta,\phi)$ is two dimensional. Let us now assume that $F$ is at least quadratic in order, containing a central minimum or maximum out of which $F$ increases or decreases in all directions.
This means Eq. \ref{eq1} defines a closed contour, which we shall call $A$, around the minimum or maximum.
If $F$ \textit{is} a quadratic function of $\theta$ and $\phi$, $A$ is an ellipse, and its area will scale linearly with the value of $\delta$ (explained in the supplementary information).
Meanwhile, the ratio of the area enclosed by $A$ to the surface area of the sphere of radius $r$  must be constant for any radius because Eq. \ref{eq1} contains no dependence on $r$.
If $r$ is increased, then in real space, as the surface area of the sphere inflates, the ellipse $A$ sweeps out an elliptical cone with its vertex at $r=0$.
Introducing radial dependence to the right-hand side of Eq. \ref{eq1}, we can modify the volume swept out by $A$ as the sphere inflates.
Since a sphere’s surface area is proportional to $r^2$, if we re-define $A$ as,
\begin{equation}\label{eq2}
    F(\theta,\phi)=\frac{\delta}{r^2},
\end{equation}
then we reshape the previously conical swept-through volume into a non-diffracting tube of constant  elliptical cross section.
Because the area of $A$ is proportional to the right-hand side of Eq. \ref{eq2}, it decreases in angular space at the same rate as $\delta/r^2$ for increasing $r$, preserving the real-space cross-sectional area of the tube which, therefore, does not diffract with distance from the centre of the sphere.
Briefly summarising, then, two features of Eq. \ref{eq2} are necessary and sufficient to immunise a closed contour to diffraction: $1/r^2$ dependence of the right-hand side, and a quadratic function $F$ of $(\theta,\phi)$ which solely increases or decreases from a central minimum or maximum in all transverse directions.
As we will show, it is possible to find non-diffracting contours in the form of Eq. \ref{eq2} in arbitrary far field radiation.
The shepherds of non-diffracting objects in the far field are radially polarised L lines, lines of pure linear polarisation, where the normally dominant transverse field components are zero.
In the far field of any radiating source, wavefronts are spherical and the prevailing polarisation components of the electric field are ‘transverse’ - the angular components comprising the field transverse to the propagation direction, $\mathbf{E}_t$, dominate, and the total field $\mathbf{E}=\mathbf{E}_t+E_r\mathbf{\hat{r}}$ lies almost orthogonal to the phase gradient.
The far field transverse components for a source with an arbitrary radiation diagram can be expressed, when sufficiently far from the source compared to its size, as,
\begin{equation}\label{eq3}
    \mathbf{E}_t=\begin{pmatrix}E_\theta \\ E_\phi
    \end{pmatrix}=\begin{pmatrix}
        E_{\theta0}(\theta,\phi)\\E_{\phi0}(\theta,\phi)
    \end{pmatrix}\frac{e^{ikr}}{r}.
\end{equation}
After the second equals sign, radial dependence has been segregated giving the purely $\theta$ and $\phi$ dependent functions $E_{\theta0}(\theta,\phi)$ and $E_{\phi0}(\theta,\phi)$.
Gauss’ law in free space $\nabla\cdot\mathbf{E}=0$ gives a differential equation which we may solve to obtain an expression for the radial component $E_r$,
\begin{equation}\label{eq4}
    E_r=\frac{const}{r^2}+\frac{i}{kr}\left(\frac{1}{\tan\theta}E_\theta+\frac{1}{\sin\theta}\frac{\partial E_\phi}{\partial\phi}+\frac{\partial E_\theta}{\partial\theta}\right).
\end{equation}
The first term $const/r^2$ accounts for a static charge (Coulomb's law) which does not radiate and may be neglected.
Then, using Eq. \ref{eq3}, we can segregate radial dependence in Eq. \ref{eq4}, obtaining,
\begin{equation}\label{eq5}
    E_r=E_{r0}(\theta,\phi)\frac{e^{ikr}}{r^2},
\end{equation}
where $E_{r0}$ is given by,
\begin{equation}\label{eq6}
    E_{r0}(\theta,\phi)=\frac{i}{k}\left(\frac{1}{\tan\theta}E_{\theta0}+\frac{1}{\sin\theta}\frac{\partial E_{\phi0}}{\partial\phi}+\frac{\partial E_{\theta0}}{\partial\theta}\right).
\end{equation}
What Eq. \ref{eq5} reveals is critical: in the far field, the radial electric field component decreases in amplitude according to $1/r^2$, while the transverse components Eq. \ref{eq3} decrease in amplitude according to $1/r$.
We now have a footing to construct a contour of the form of Eq. \ref{eq2} in the intensity of the transverse and radial field components.
Calculating $\mathbf{E}^*_t\cdot\mathbf{E}_t=|E_\theta|^2+|E_\phi|^2$ and equating to $|E_r|^2$ with segregated radial dependence, we can write an expression defining a line on the sphere of radius $r$,
\begin{equation}\label{eq7}
    \frac{1}{r^2}|E_{\theta0}|^2+\frac{1}{r^2}|E_{\phi0}|^2=\frac{\delta}{r^4}|E_{r0}|^2,
\end{equation}
where $\delta$ is a constant, defined as the ratio between transverse and longitudinal field intensity.
Simplifying, we arrive at Eq. \ref{eq8} below,
\begin{equation}\label{eq8}
    \frac{|E_{\theta0}(\theta,\phi)|^2+|E_{\phi0}(\theta,\phi)|^2}{|E_{r0}(\theta,\phi)|^2}=\frac{\delta}{r^2},
\end{equation}
which immediately satisfies the first criterion for a diffraction free contour of the form of Eq. \ref{eq2} – a $1/r^2$ dependence of the right-hand side.
We may satisfy the second criterion if a zero in the transverse field (a radially polarised L line) is placed at an angular position $(\theta_1,\phi_1)$.
Because intensity is always greater than or equal to zero, a minimum in the transverse field intensity is created and Eq. \ref{eq8} must organise a closed loop in angular space around $(\theta_1,\phi_1)$.
Moreover, in the close neighbourhood of the transverse field zero, a lowest order series expansion in $(\theta,\phi)$ means that the approximate transverse components $E_{\theta0}\approx\tilde{E}_{\theta0}$ and $E_{\phi0}\approx\tilde{E}_{\phi0}$ behave linearly (crossing zero at $(\theta_1,\phi_1)$) while the radial component is approximately constant, equal to $E_{r0}(\theta_1,\phi_1)\neq0$, provided the electric field Jacobian matrix is non-zero.
It follows that the transverse field intensity $\mathbf{E}^*_t\cdot\mathbf{E}_t$ varies quadratically with $(\theta,\phi)$ and Eq. \ref{eq8}, under this first order approximation, defines an ellipse.
Finally, re-writing Eq. \ref{eq8} explicitly under our approximation,
\begin{equation}\label{eq9}
    \frac{|\tilde{E}_{\theta0}|^2+|\tilde{E}_{\phi0}|^2}{|E_{r0}(\theta_1,\phi_1)|^2}=\frac{\delta}{r^2},
\end{equation}
both requirements for a non-diffracting closed contour have been satisfied.
If $\delta=1$, then this contour encloses the region around the L line where electric spin has a dominating transverse component.
In summary, if a zero exists at some position $(\theta_1,\phi_1)$ in the transverse field components of arbitrary far field radiation, then the transverse field intensity $\mathbf{E}^*_t\cdot\mathbf{E}_t$ must cross equality with the radial field intensity $|E_r|^2$ (or $\delta|E_r|^2$, multiplied by a small constant) on an ellipse containing $(\theta_1,\phi_1)$.
When propagated in the radial direction, this ellipse draws out a non-diffracting tube of constant cross section.

\subsection{Tube width}
Although the real-space cross-section of the non-diffracting tube is invariant with distance, its constant width and ellipticity depend on the radiation diagram of the source.
The confinement of the transverse field zero, and therefore the diameter of the tube, depends on the first-order derivatives of $E_{\theta0}$ and $E_{\phi0}$ in the position of the zero, which are neatly organised in the transverse field's Jacobian matrix,
\begin{equation}
    \mathbf{J}_0=\begin{pmatrix}
        \frac{\partial E_{\theta 0}}{\partial \theta}&\frac{\partial E_{\theta 0}}{\partial \phi}\\
        \frac{\partial E_{\phi 0}}{\partial \theta}&\frac{\partial E_{\phi 0}}{\partial \phi}\\
    \end{pmatrix}.
\end{equation}
The area of the tube cross section (an ellipse) is given by $A=\pi\delta \sin{\theta} |E_{r0}|^2/|\det{\mathbf{J}_0}|$, where $E_{r0}$ and $\mathbf{J}_0$ are evaluated in the position of the transverse field zero.
A large value of $|\det{\mathbf{J}_0}|$ corresponds to a narrower non-diffracting tube, and vice versa, while $\sin\theta$ accounts for local distortion of the $\theta,\phi$ axes over the sphere surface.
Generally, the elliptical tube cross-section has semi-axes on the order of a (significant) fraction of a wavelength depending, of course, on the choice of $\delta$ in Eq. \ref{eq9} (the semi-axes scale linearly with $\sqrt{\delta}$).
For the simple case of the axial transverse field zero of a linearly polarised dipole, radiating according to $E_\theta=(1/r)\exp(ikr)E_0\sin\theta$ and $E_\phi=0$, it may be shown that the (cylindrical) non-diffracting tube around the zero has a radius of $\sqrt{\delta}\lambda/\pi$.
Further details are found in the SI.

\subsection{C lines}
Next, we address the lines of circular polarisation shown in Fig. \ref{fig1} which lie parallel to the non-diffracting intensity-ratio tube and maintain their proximity with $r$.
This is another imprint created by the transverse field zero, and can be explained by the domination of the transverse components in the far field, that is, $\mathbf{E}\approx\mathbf{E}_t$.
C lines are phase singularities in the complex scalar field $\psi$, existing where $\psi=\mathbf{E\cdot\mathbf{E}}=0$.
A C line is therefore an intersection of the (in 3D space) surfaces $Re\{\psi\}=0$ and $Im\{\psi\}=0$, which we will call $M$ and $N$ respectively.
Under the condition $\mathbf{E}\approx\mathbf{E}_t$, $M$ and $N$ are brought near to each other by a zero in the transverse field at $(\theta_1,\phi_1)$, although they will not intersect at $(\theta_1,\phi_1)$ due to the non-zero radial component.
However, just like the field intensity, $\psi$ has an approximately quadratic region near to $(\theta_1,\phi_1)$ where, due to their natural proximity, it is likely that $M$ and $N$ intersect in multiple locations.
Since the transverse field zero is preserved radially, the surfaces $M$ and $N$ will always remain near to each other in the radial direction and, with more than one intersection, produce parallel radial C lines close to the non-diffracting tube of transverse and radial intensity.
An interesting characteristic of the transverse field zero comes from the fact that the linear approximation of the transverse components in the neighbourhood of $(\theta_1,\phi_1)$ becomes increasingly better with the radial distance from the source, because for a given real-space distance, the angular distance gets smaller.
This leads to a deepening symmetry of the $Re\{\psi\}=0$ and $Im\{\psi\}=0$ surfaces in the region of $(\theta_1,\phi_1)$, becoming almost exactly quadratic and limited to an even number of intersections.
In fact, the surfaces $Re\{\psi\}=0$ and $Im\{\psi\}=0$ assume one of three forms on the sphere of radius $r$: an ellipse, a typical hyperbola or a degenerate hyperbola, any two of which will intersect in an even number of locations (a more detailed picture is provided in the supplementary information).
An odd number of C lines may briefly accompany the non-diffracting intensity tube in the intermediate field region before one (or more, in an odd number) C line diverges, leaving an even number of C lines to propagate into the far field.
The number of associated non-diverging C lines could define a topological index for the vector field zero $\mathbf{E}_t=\mathbf{0}$ (similarly to the indices of three-component zeros $\mathbf{E=0}$ in \cite{Vernon2023}), in parallel with the topological charge $\pm1$ of the complex scalar zeros $E_\theta=0$ and $E_\phi=0$.

\subsection{Stability}
In this article we described electric fields using complex phasors meaning a zero in the transverse field $\mathbf{E}_t$ [Eq. (\ref{eq3})] is time-independent.
Such time-independent zeros are codimension 4 objects, unstable to perturbation and do not occur naturally in interference like other paraxial point singularities (i.e., C points \cite{Dennis2002}).
Sensitivity to noise is a challenge for practical application but could be alleviated if an extra parameter of the source, for example wavelength, is adjustable.
This way, perturbed zeros and their non-diffracting structures can be recovered by re-tuning the source wavelength \cite{Spaegele2023}.
The non-diffracting phenomenon can also be identified in any source's instantaneous electric field, which is described by a real, time-dependent vector $\boldsymbol{\mathcal{E}}=Re\{\mathbf{E}\exp{-i\omega t}\}$.
Instantaneous zeros, although not fixed in time and position, appear twice per oscillation at points of linear polarisation just as the electric field vector changes direction.
That the basic $1/r^2$ and $1/r$ dependency applies both to the real and imaginary parts of the radial and transverse field phasors respectively produces the same non-diffracting tube structures in the time-varying electric field vector length.
This is straightforward to show by introducing time-dependence [$\exp(-i\omega t)$] to Eqs. (\ref{eq3})-(\ref{eq6}), taking the real part, and computing the square length of these instantaneous components (see supplementary information).
We may conclude from this fact and topological arguments that, at any given instant, non-diffracting structures appear indiscriminately in all far field radiation.
According to the hairy ball theorem, the electric field lines at an instant in time cannot be `combed' over a sphere surface without incurring at least one singularity where the direction of the field lines is undefined and the transverse field vanishes.
For these reasons, non-diffracting structures which always accompany instantaneous transverse field zeros are completely fundamental phenomena in light radiated by localised sources.
Further demonstration of the ubiquity of non-diffractive behaviour is given in the supplementary information.


\section{Radiation from electromagnetic multipoles}		\label{sec:two}	

So far, we have argued that any transverse field zero, propagating an unrestricted distance into the far field of any source, naturally retains a small neighbourhood where light is non-paraxial.
Inside this non-paraxial region is a non-diffracting elliptical tube defined where the transverse and radial field intensities are equal (or proportional by a small constant), and an even number of parallel, non-diverging C lines.
These are the features which always occur near the transverse zero of a generic source, however, more symmetric sources can produce degenerate non-diffracting structures such as C surfaces.
To show this, we now consider the radiation from point-like multipoles.
Following the textbook formalism \cite{jackson1999classical}, we express the electric field of electric-type multipole radiation in terms of spherical harmonics $\mathrm{Y}_{l,m}(\theta,\phi)$,
\begin{eqnarray}
\label{eq:sphercomp}
E_{\theta}&=& \frac{1}{2}\Big[ \sqrt{(l-m)(l+m+1)} \mathrm{Y}_{l,m+1}(\theta,\phi)e^{-i\phi}-\sqrt{(l+m)(l-m+1)} \mathrm{Y}_{l,m-1}(\theta,\phi)e^{i\phi}\Big]\frac{E_0e^{ikr}}{kr},\nonumber \\ 
E_{\phi} &=& \frac{im}{\sin\theta}\mathrm{Y}_{l,m}(\theta,\phi)\frac{E_0e^{ikr}}{kr}, \\
E_r&=&-il(l+1)\mathrm{Y}_{l,m}(\theta,\phi)\frac{E_0e^{ikr}}{(kr)^2},\nonumber
\end{eqnarray}
where $E_0$ is a normalization constant.
Magnetic-type multipole radiation has a zero radial component of electric field by construction, and it is not considered here.

As follows from Eq.(\ref{eq:sphercomp}), the radial component of the field falls off with the distance to the source by a factor of $(kr)$ faster than transverse field components.
It can also be seen that the $\phi$-components and $r$-components turn to zero at the same angles, which is not the case for zeros of the $\theta$-components. 

Let us consider multipole radiation with $m=0$ that corresponds to axially-symmetric multipoles (multipoles for which $m\neq0$ are not relevant as they do not contain time-independent zeros in the transverse field).
In this case $E_\phi=0$ and we look for values of $r(\theta)$ that satisfy the C line condition $\mathbf{E\cdot\mathbf{E}}=0$.
First, we identify zeros of the $E_\theta$ field component at $\theta=0\ \mathrm{and}\ \pi$ (on the $z$-axis).
Since the $E_r$ component is nonzero, we obtain an L-type polarisation singularity on the $z$-axis.
After Taylor-expanding the spherical harmonics around $\theta=0\ \mathrm{and}\ \pi$, we arrive at the C line condition,
\begin{eqnarray}
\label{eq:C-cond}
    &&kr\theta=2, \nonumber\\
    &&kr(\pi-\theta)=2,
\end{eqnarray}
{\it independently} of the value of $l\geq 1$. Recognizing that the quantity $r\theta,r(\pi-\theta) =r_\perp$ is the transverse distance from the $z-$axis, it follows that the C line is parallel to the $z-$axis, separated from it by the distance $r_\perp=\lambda/\pi$, where $\lambda$ is the wavelength; hence C lines appear parallel and $2\lambda/\pi$ apart.  Since the solution is valid for any value of $\phi$, the C lines become a C surface in three dimensions, which is a cylinder of diameter $2\lambda/\pi$. This cylinder is tapered toward small $r$ near the origin, where the radial field dominates. For an electric dipole, this solution was discovered in Ref.\cite{Mok2022}. Here we extend it to a general case of arbitrary multipoles and to both C-type and L-type polarisation singularities.

For higher multipoles $l>1$, additional solutions for C lines are found near zeros of $E_\theta$ defined by the zeros of spherical harmonics $\mathrm{Y}_{l,\pm 1}(\theta,\phi)$: $\theta_0=90^o$ for $l=2$, $\theta_0=63.4^o,116.6^o$ for $l=3$, $\theta_0=49.1^o, 90^o, 130.9^o$ for $l=4$, $etc.$ (see, $e.g$, Ref.\cite{varshalovich1988quantum}).
Note that for these angles the separation distance between C lines is different from the $\theta_0=0,\pi$ case---it is found to be exactly a factor of two smaller and equal $\lambda/\pi$---but one important feature remains the same: the C lines extend to infinite $r$, while staying parallel to the line $\theta=\theta_0$.
Due to axial symmetry, in three dimensions the C lines for $\theta\neq 0,\ \theta\neq\pi$ turn into pairs of parallel conical surfaces.
Since for the transverse field zeros only the radial $E_r$ component of the field is nonzero, these angles define $r$-polarised L lines for $\theta_0=0,\pi$, and L cones for all other values of $\theta_0$.
It should also be noted that in a general case of superposition of several axially symmetric multipoles with different values of $l$, the L line and a central C cylinder remain unchanged.
In addition, for even-valued multipoles ($l=2,4,...$), a horizontal L plane is present, sandwiched between two C planes. 

Circular polarisation of the electromagnetic radiation is associated with a (pseudo)vector $\mathbf{s}$ of spin polarisation (see, for example, Ref.\cite{afanasev2023nondiffractive}), by definition, 
\begin{equation}
\label{eq:spinvec}
    |E|^2 \mathbf{s}=i\mathbf{E}\times\mathbf{E^*}.
\end{equation}
For nonzero fields, the C line condition $\mathbf{E\cdot\mathbf{E}}=0$ is equivalent to the condition $|\mathbf{s}|=1$.
Using the field components from Eq.(\ref{eq:sphercomp}), it is straightforward to show that on the C surfaces, the azimuthal component of the polarisation vector $\mathbf{s}$ reaches unity, $s_\phi=\pm1$, while the components $s_r,s_\theta$ are zero.

\begin{figure}[t]
    \centering
    \includegraphics[width=\textwidth]{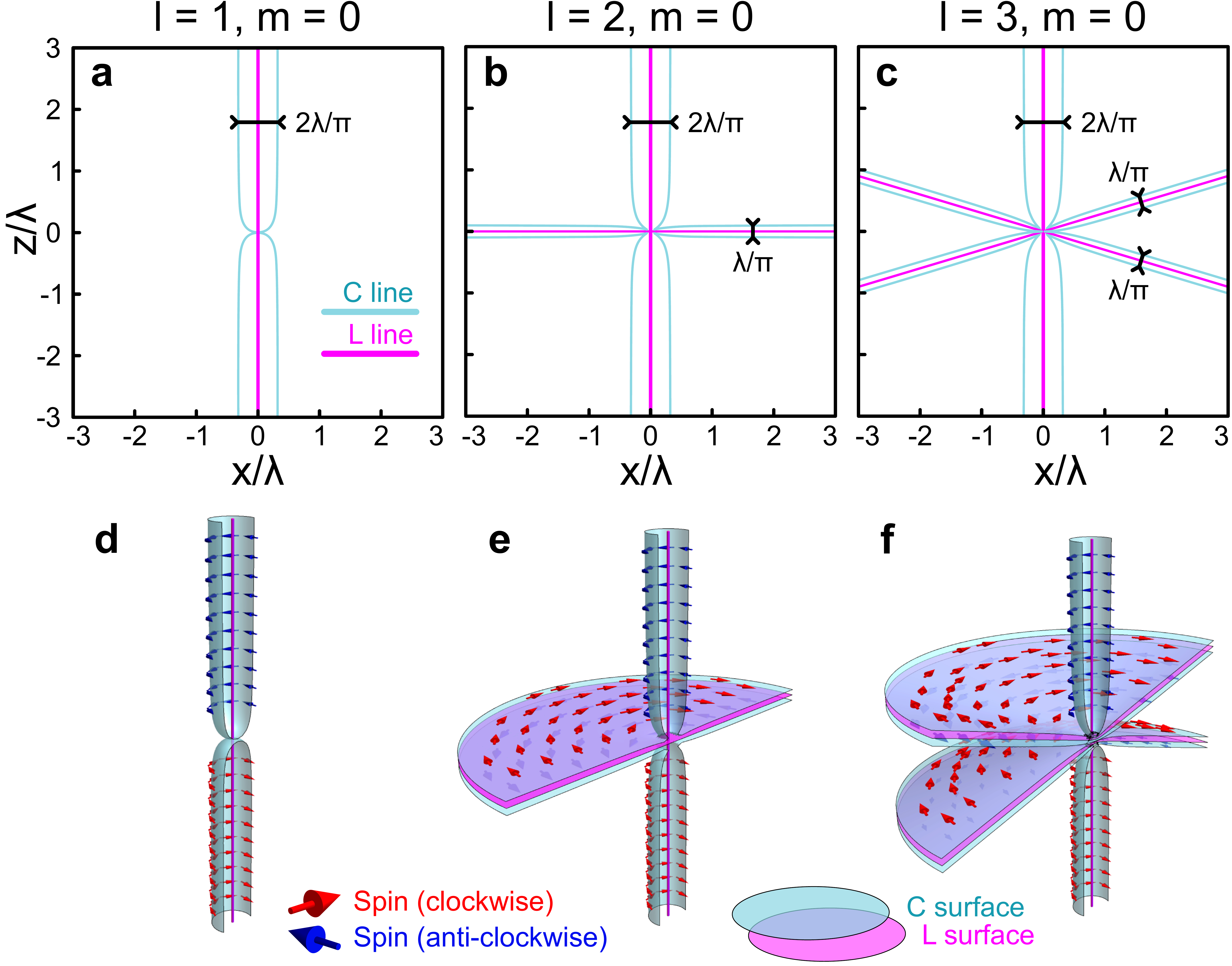}
    \caption{A real-space map of polarisation singularities, with (xz)-plane cut (a-c) showing C lines (blue) and L lines (magenta) of electric field of axially-symmetric multipole radiation.
    The point-like radiation source is located in the origin.
    The distances are in units of wavelength.
    The multipoles have $m=0$, $l=1$ ([a, d] dipole), $l=2$ ([b, e] quadrupole) and $l=3$ ([c, f] sextupole).
    The bottom row (d-f) shows 3D-counterparts of the upper plots, together with the directions of spin polarisation {\bf s} on C surfaces.
    Red arrows indicate a clockwise direction and blue arrows are counter-clockwise; only $s_\phi$ component is nonzero.}
    \label{fig:CL-2D}
\end{figure}

The above results are illustrated in the following Figures.
We show L and C lines in the $l=1,2,3$ multipole radiation in Figure~\ref{fig:CL-2D} in position space over a region of a few wavelengths.
One can see central L lines (magenta) sandwiched between pairs of C lines (blue) that become parallel after $r\gtrsim \lambda/4$ and extend to infinite distances from the source; three-dimensional versions of the same plots also indicate directions of polarisation vector {\bf s}.
One can see the L line along the $z$-axis, and C lines turning into a cylinder in 3D; for the angles away from the $z$-axis, L lines in 3D turn into conical L surfaces sandwiched between pairs of conical C surfaces.
Neither the cylinder's radius, nor separation distance between conical C surfaces increase with the distance from the source in the far field, demonstrating the non-diffractive features discussed above. 
Let us consider the direction of spin polarisation $\mathbf{s}$ of the field, as defined by Eq.~(\ref{eq:spinvec}).
As shown in Figure \ref{fig:CL-2D}, only the $s_\phi$ component of polarisation survives in the axially symmetric case, with counter-clockwise (clockwise) circulation on the C cylinder for positive (negative) $z$.
For $l=2$, the upper C surface at $\theta=\pi/2$ shows clockwise circulation, while the lower one has counter-clockwise circulation.
This feature continues for higher multipoles: the pairs of parallel C surfaces have opposite directions of the normalised spin $\mathbf{s}$.
Starting from wavelength-scale distances away from the source, the circulation of the polarisation vector around the cylindrical C surface is calculated to be,
\begin{equation}
\label{eq:spincirc}
    \oint \mathbf{s}d\mathbf{l}=2\lambda (z\gtrsim \lambda); -2\lambda (z\lesssim -\lambda).
\end{equation}

Thus, the multipole radiation from axially symmetric multipoles ($m=0$) reveals non-diffractive C surfaces near zeros of the transverse field, as was concluded from general arguments in Section \ref{sec:one}.
A cylindrical C surface of constant diameter $2\lambda/\pi$ is wrapped around the $z$-axis, while pairs of parallel conical C surfaces accompany L cones, with a number of cones increasing for higher multipoles.
The same arguments apply to magnetic fields radiated from magnetic multipoles, which also have a radial field component that falls off by a factor of $1/kr$ faster than the transverse ones.

\begin{figure}[t]
    \centering
    \includegraphics[width=\textwidth]{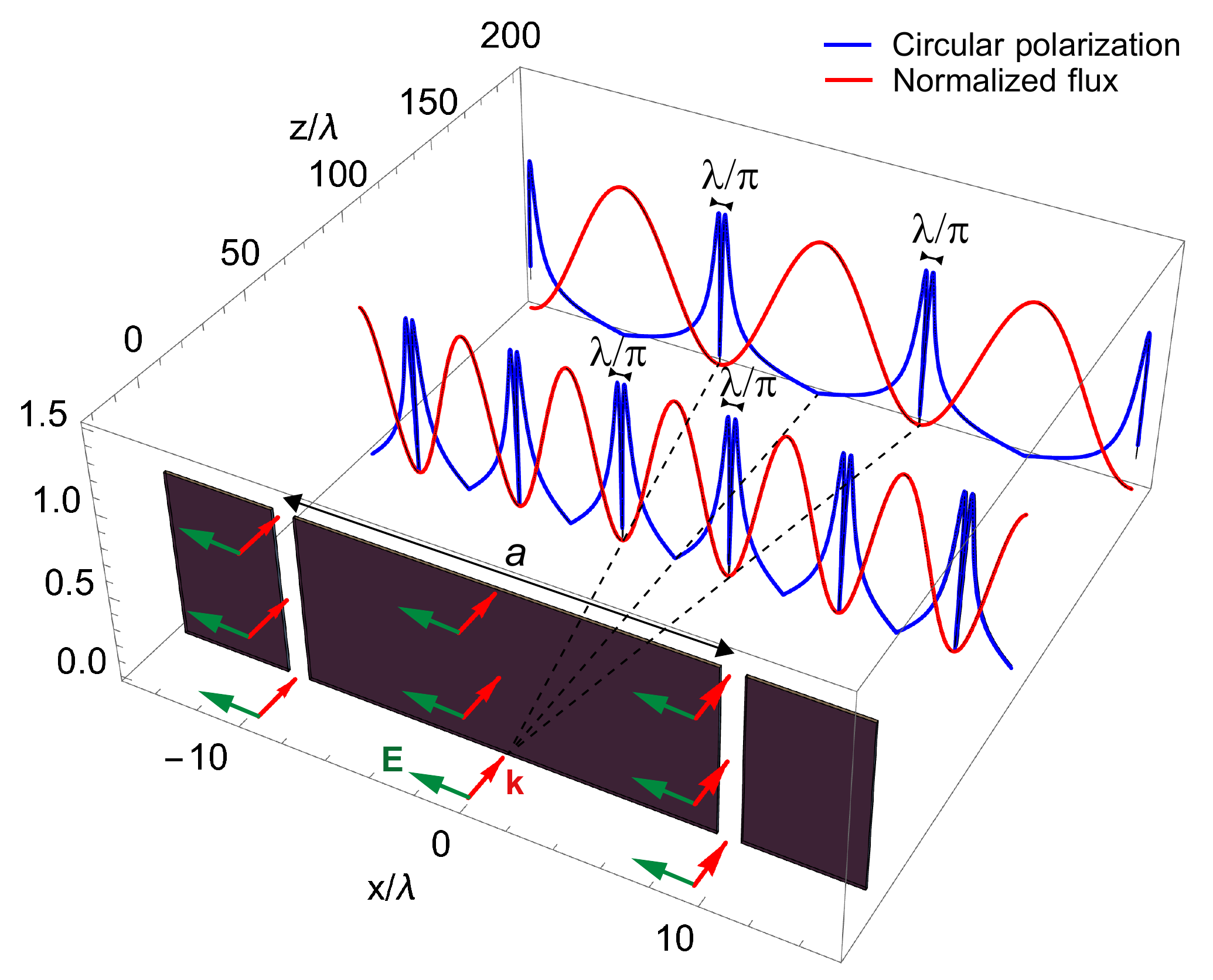}
    \caption{
    Intensity (red curves) and circular polarization $|\mathbf s|$ (blue curves) for Young's double-slit interference. The incident light is linearly polarized perpendicular to the slits. Pairs of C-lines are developed due to interplay between small but nonzero $z$-components of the electric field normal to the screen and (otherwise dominant) in-plane $x$-components that turn to zero due to destructive interference between two slits. For each C-line pair, photon polarization vectors point along $y$-axis, reaching unit length, and are opposite in sign: $s_y=\pm1$.
    }
    \label{fig:DSlit}
\end{figure}

\section{Double-slit experiment} \label{sec:doubleslit}

Non-diffracting polarisation profiles are widespread in light because of the fundamental distinction between the radial dependence of the transverse ($1/r$) and longitudinal ($1/r^2$) electric field components, as shown in Eq. (\ref{eq3}) and (\ref{eq5}).
Although longitudinal far field components are typically ignored, they cannot be neglected near to intensity minima and, when considered, reveal that non-diffracting polarisation singularities have been hiding in even the most elementary and well-studied of optics experiments.
Let us consider, for example, the celebrated double-slit experiment by Thomas Young performed in the 19th century to demonstrate the wave nature of visible light.
Here, in addition to standard textbook analysis, we will not neglect small components of the electric field oriented normal to the screen. 
The experiment's geometry is shown in Fig. \ref{fig:DSlit}.
Two narrow parallel slits aligned along the $y$-direction are separated by a distance $a$ in the $x$-direction.
A projection screen is located at a distance $z$ in the $(xy)$-plane.
Dropping overall factors, the electric field at a position $x$ on the projection screen can be expressed as superposition of coherent light from the individual slits,
\begin{equation}\label{eq:superposition}
    \mathbf{E}=\mathbf{E}_1+\mathbf{E}_2=\mathbf{E}_{10} e^{ikr_1}+\mathbf{E}_{20}e^{ikr_2}, 
\end{equation}
where $r_{1,2}$ define the light's path from each slit and $|\mathbf{E}_{10}|=|\mathbf{E}_{20}|=E_0$.
The field component that lies in the screen plane is dominant and equal to $E_x=E_0(e^{ikr_1}+e^{ikr_2})$, assuming near-parallel rays: $x/z\ll 1, a/z\ll 1$.
The difference in light's path $r_2-r_1=2\pi m/k$ results in constructive interference and $r_2-r_1=\pi(2m+1)/k$ results in destructive interference, setting the angular positions of the interference fringes to $\alpha \equiv x/z=\lambda m/a$ and $\lambda (2m+1)/2a$ for maxima and minima, respectively, where $m$ is an integer.
Let us now consider linearly polarized light in $x$-direction incident on the slits, and include the often neglected smaller components of the field normal to the screen.
In a small-angle approximation, $\sin\alpha\approx\alpha$, we have $E_z=-E_0(\alpha_1 e^{ikr_1}+\alpha_2 e^{ikr_2})$. In this approximation, $E_z$ is simple to understand as a result of the parallax angle of each point in the screen relative to the two slits. 
The angular positions of individual slits are $\alpha_{1,2}\approx(x\mp a/2)/z$. 
Thus, we obtain,
\begin{equation}
    E_z=-E_0\left[\frac{x}{z}(e^{ikr_1}+e^{ikr_2})-\frac{a}{2z}(e^{ikr_1}- e^{ikr_2})\right],
\end{equation}
and the following expression for the ratio of field components:
\begin{equation}
\label{eq:ExEz}   
 \frac{E_z}{E_x}=-\frac{x}{z}+\frac{a}{2z}\frac{1- e^{ik(r_2-r_1)}}{1+e^{ik(r_2-r_1)}}.
\end{equation}
Expanding the exponential functions in Eq. (\ref{eq:ExEz}) near locations of the interference minima of the $E_x$ component, $e^{ik(r_2-r_1)}\approx-1+ik\delta r$, we see that the $E_z$ component remains nonzero at the dark fringe, with a ratio $E_z/E_x\to \pm i\infty$ as $\delta r\to\mp 0$.
It also follows, after some algebra, that $E_z/E_x=\mp i$ for the optical path's offset from the interference minima by $\delta r=\pm a/(z k)$ that corresponds to $x$-positions on the projection screen offset by $\delta x=\pm 1/k =\pm\lambda/2\pi$.
Thus, for each interference minimum we obtain a pair of C lines with opposite polarization for the predicted $x$-positions.
These C lines are separated by $\lambda/\pi$ on the projection screen along the $x$-direction {\it independently of the distance $z$ to the slits' plane}, as illustrated in Fig. \ref{fig:DSlit}. While the interference intensity pattern diffracts, with minima appearing at fixed angular positions, the accompanying C lines remain parallel in real space, immune to diffraction, within the ever-spreading intensity minimum. It is straightforward to reproduce the same result following the formalism of Section 2, by applying Gauss’ law in free space $\nabla\cdot\mathbf{E}=0$ and a paraxiality condition $\partial{E_z}/\partial{z}=i k E_z$.
We note here that for the light polarized along the $y$-direction (along the slits), the polarization on the projection screen remains linear, since longitudinal field components would not arise (but the effect would instead be present in the magnetic field).


\section{Discussion}\label{sec:conclusions}
We studied polarisation singularities of the electromagnetic fields in a general field formalism for radiation from arbitrary point-like sources, delivering a complete generalisation of the non-diffracting intensity ratio feature found in vector vortex beams \cite{afanasev2023nondiffractive}.
In the far field region of any localised source, the transverse $\boldsymbol{\hat{\theta}}$ and $\boldsymbol{\hat{\phi}}$ field components dominate such that any radial component is neglected.
However, when both transverse field components are zero (in a `transverse field zero'), the radial component becomes dominant and cannot be ignored.
The transverse field zero is, in fact, a radially polarised L line which carries with it a small neighbourhood where light has non-paraxial characteristics, even at far field distances.
Generically this non-paraxial region contains an even number of C lines, and intensity ratio contours which can be defined around the zero where the transverse and radial field intensities are proportional by a
constant.
Remarkably, these features do not diffract: the C line pairs do not diverge and the intensity ratio contours define a tube in 3D which has constant cross section.
Time-varying transverse field zeros are stable singularities, and a minimum of one and its coupled non-diffracting structures are always present in far field radiation.
Instantaneous non-diffracting phenomena are thus produced by all physical sources of propagating light.

These arguments were illustrated with examples of multipole radiation and double slit interference.
Our conclusions are general to any wavelength, which influences only the real-space transverse dimension of non-diffracting polarisation features, and the polarimetry techniques required to measure them.
A straightforward method of reconstructing such polarisation structures is directional interference with a sub-wavelength probe.
As previously demonstrated \cite{bauer2015}, collection of transmitted and scattered light from a probe scanned near transverse field zeros can enable one to experimentally extract spatial field components and derive the non-diffractive intensity-ratio tube.
Moreover, a sub-wavelength probe placed in the parallel C lines or C surfaces, which are separated by a significant fraction of the wavelength, will scatter circularly polarised light detectable by a polarimeter.
As demonstrated in the supplementary information, polarisation structures generated by antenna arrays in the RF regime have a substantial dimension that is easily identifiable through these techniques.
Diffraction-independent spatial separation of polarization singularities demonstrated in our paper are, at first glance, in conflict with uncertainty relations, since corresponding angular separation would be inversely proportional to the the distance to the source.
However, since this property is a feature of dark fringes in the interference patterns of light, this is consistent with existing approaches to super-resolution microscopy.
We are hopeful that our observations will help advancing super-resolution microscopy enhanced with polarization features, radar sensing techniques, and chiral-dependent light probes of matter.



\section{Acknowledgements}
A. A. acknowledges support from US Army Research Office Award No. W911NF-23-1-0085.
A. J. V. is supported by EPSRC Grant EP/R513064/1.

\section{Data availability}
Data underlying the main results presented in this paper can be computed from the simple equations provided, so no data was generated for the results.
To generate the fields in the illustrative Fig.~1, a source was modelled as a collection of ten randomly spaced electric dipoles which were polarised deliberately to create a transverse field zero. Data for Fig.~1 may be obtained from the authors upon reasonable request.

\section{Competing Interests}
The Authors declare no competing interests.
\end{document}